\documentclass[a4paper,11pt]{article}

\usepackage{amsthm}
\usepackage{amsmath,amssymb,latexsym,amsfonts,mathrsfs}
\usepackage[dvipdfmx]{graphicx}

\usepackage{color}

\usepackage{fancyhdr}

\newcommand{\nn}{\nonumber}

\newcommand{\SCR}[1]{{\mathscr #1}}

\newcommand{\J}[1]{\left\langle #1 \right\rangle}

 \usepackage{amsthm}

 \newtheorem{Thm}{Theorem} 
  
\newtheorem{Cor}[Thm]{{Corollary}}
\newtheorem{Ass}[Thm]{{Assumption}}

\newtheorem{Rem}[Thm]{{Remark}}

\begin{document}

\begin{center}
{\bf \Large Energy Stableness for Schr\"{o}dinger Operators with Time-Dependent Potentials 
 } 
 \end{center} 
 \begin{flushleft}
by {\bf \large Masaki Kawamoto} \\ 
 Department of Engineering for Production, Graduate School of Science and Engineering, Ehime University, 3 Bunkyo-cho Matsuyama, Ehime, 790-8577. Japan. \\ 
Email: kawamoto.masaki.zs@ehime-u.ac.jp
\end{flushleft}

\begin{center}
\begin{minipage}[c]{300pt}
{\bf Abstract}. {\small 
In this paper, we prove the energy stable property for time-dependent (generalized) Schr\"{o}dinger operators by using Hardy inequality. Such property acts very important roles in quantum scattering theory and nonlinear problem. As an application, we prove Sobolev type inequality.
}
\end{minipage}
\end{center}

\begin{flushleft}
\end{flushleft}
\section{Introduction}
We consider the (generalized) Schr\"{o}dinger equations with time-dependent potential; 
\begin{align*}
i \partial _t \phi(t,x) &= ((- \Delta )^{\theta} /(2m)+ V(t)) \phi(t,x), \\ 
\phi (s,x) &=\phi (s) \in H^1({\bf R}^n), 
\end{align*}
where $\theta \geq 1/2$, $x=(x_1,x_2,...,x_n) \in {\bf R}^n$, $\Delta = \partial _1^2 + \partial _2^2 + ... + \partial _n^2$ is the Laplacian, $m>0$, $s \in {\bf R}$ is a fixed cnnstant, and  $V(t)$ is a potential; a multiplication operator of real valued function $V(t,x)$ defined later. We use notations $p= -i \nabla = -i(\partial _1, \partial _2, ..., \partial _n)$ i.e.,  $-\Delta = p^2$. Let $H(t) = p^{2 \theta}/(2m) +V(t)$ and call {\em energy} in $t$. We say a family of unitary operators $\{U(t,s) \}_{(t,s) \in{\bf R}^2}$ a propagator for $H(t)$ if each component satisfies 
\begin{align*}
& i \frac{\partial}{\partial t}U(t,s) =H(t) U(t,s) , \quad i \frac{\partial}{\partial s}U(t,s) = - U(t,s) H(s) , \\ & U(t,\tau) U(\tau,s) = U(t,s), \quad U(s,s) = \mathrm{Id}_{L^2({\bf R}^n)}. 
\end{align*} 
In this paper, we state the following assumptions on the potential; 
\begin{Ass}\label{Ass1}
Let $V (t,x) $ : ${\bf R} \times {\bf R}^n \to {\bf R}$ satisfies $ V \in C^1({\bf R} \, ; \, C^1({\bf R}^n))$. Suppose that for all $u \in H^{\theta} ({\bf R}^n)$, there exist positive constants $C_{H} >0$ and $C>0$ such that 
\begin{align} \label{2}
(p^{2 \theta} u,u)_{L^2({\bf R}^n)} \geq C_{H} ( V(t) u, u )_{L^2({\bf R}^n)}
\end{align}
and 
\begin{align} \label{1} 
\int_s^{\infty} \left\| \frac{\partial}{\partial t} V(t, \cdot) \right\|_{L^{\infty}({\bf R}^n)} dt \leq C .
\end{align}
holds. Moreover, assume that the propagator $U(t,s)$ uniquely exists and that for all $t \in {\bf R}$, $U(t,s)$ satisfies 
\begin{align*}
U(t,s) H^{\theta}({\bf R}^n) \subset H^{\theta}({\bf R}^n) .
\end{align*} 
\end{Ass}
In order to obtain energy stableness, we further state the following two conditions; 
\begin{Ass} \label{Ass2}
For all $u \in H^{\theta}({\bf R}^n)$, potential $V(t)$ satisfies the following inequality 
\begin{align*}
\left( \left( p^{2\theta}/4m + V(s) - \int_s^{t} \left\| \frac{\partial}{\partial {\tau}} V(\tau, \cdot ) \right\|_{L^{\infty} ({\bf R}^n)}d \tau \right)u,u \right)_{L^2({\bf R}^n)} \geq 0 . 
\end{align*} 
\end{Ass} 
\begin{Ass} \label{Ass3}
Energy $H(t)$ is strictly positive, that is, for all $t \in {\bf R}$, $x \in {\bf R}^n$ and $u \in H^{\theta}({\bf R}^n)$, there exists $\delta >0$ such that 
\begin{align} \label{9} 
(H(t)u,u)_{L^2({\bf R}^n)} = \left( \left( p^{2 \theta}/(2m) +V(t,x) \right) u ,u \right)_{L^2({\bf R}^n)}\geq \delta \left( p^{2 \theta} u,u \right)_{L^2({\bf R}^n)}
\end{align} 
\end{Ass}
\begin{Rem}
If the potential is independent of the time, the assumption \ref{Ass1} admits {\em Coulomb type potential}, that is, $V= C_{H,n} |x|^{-2 \theta_0} $, where $0< \theta _0 < \theta$ and $C_{H,n} >0$ depends only on the dimension $n \in {\bf N}$, see e.g. Secchi-Smets-Willem \cite{SSW} since we can obtain the unique existence of the propagator $U(t,s) = e^{-i(t-s) H}$ with $H= p^{\theta}/(2m) +V$ by using the Stone's theorem and self-adjointness of $H$. On the other hand, in the case where potential depends on time, it seems difficult to include singular potentials such as Coulomb type potential in Assumption \ref{Ass1} because of \eqref{1}. For $\theta =1$ the unique existence of the propagator is guaranteed by Yajima \cite{Ya} even if the potential depends on time and has singularities. If $V$ is bounded, then we can easily prove the unique existence of $U(t,s)$.
\end{Rem}
\begin{Rem}
 As an example, we consider the potential written as the form
\begin{align*}
V(t,x) = V_0(x + c(t)) \in {\bf R}, 
\end{align*}  
where $ V_0 \in \SCR{B}^1({\bf R})$, and $c=(c_1,...,c_n) $ satisfies  that for all $j \in \{ 1,...,n\}$, $c_j \in \SCR{B}^1({\bf R})$ and integrable condition
\begin{align*}
\int_s^{\infty} | c'(t) | dt < \infty . 
\end{align*}
Additionally if $V(t)$ satisfies
\begin{align*}
\sup_{x \in{\bf R}^n} \left| |{x}|^{2 \theta_0 } V(t,x) \right| \leq C_H
\end{align*}
for some constant $C>0$ and some $\theta _0 > \theta$ then $V$ satisfies Assumption \ref{Ass1}. For $\theta =1$, such a potential appears for Schr\"{o}dinger equations with time-decaying electric fields, see e.g., Adachi-Fujiwara-Ishida \cite{AFI}. What we emphasize here is we do not need to assume that $|c(t)| \to 0$ as $|t| \to \infty$.  
\end{Rem}
Under the assumption \ref{Ass1}, we have the following Theorem; 
\begin{Thm}[Stableness for Laplacian]\label{T1}
Let $\phi(s) \in H^{\theta}({\bf R}^n)$ satisfies that for some $0<a < R $, $\mathrm{supp} (\hat{\phi}(s)) \subset \left\{ \xi \in {\bf R}^n \, | \, a \leq |\xi| \leq R \right\}$, where $\hat{\cdot}$ stands for the Fourier transform. If we assume Assumption \ref{Ass1} and \ref{Ass2}. Then for any $t \in {\bf R}$ there exists a $t-$independent constant $\tilde{a} >0$ such that 
\begin{align} \label{3}
\left( p^{2 \theta} U(t,s) \phi(s) , U(t,s) \phi(s) \right)_{L^2({\bf R}^n)}  \geq \tilde{a} \left\| 
\phi(s)
\right\|^2_{L^2({\bf R}^n)}
\end{align}
holds. On the other hand, if we assume Assumption \ref{Ass1} and \ref{Ass3}. Then for any $t \in {\bf R}$ there exists a $t-$independent constant $\tilde{R} >0$ such that 
\begin{align} \label{4}
\left( p^{2 \theta} U(t,s) \phi(s) , U(t,s) \phi(s) \right)_{L^2({\bf R}^n)}  \leq \tilde{R} \left\| 
\phi(s)
\right\|^2_{L^2({\bf R}^n)}
\end{align}
holds. Hence if we assume Assumption \ref{Ass1}, \ref{Ass2} and \ref{Ass3}, then for any $t \in {\bf R}$ there exist $t-$independent constants $\tilde{a} >0$ and $\tilde{R} >0$ such that 
\begin{align*}
\tilde{a} \left\| 
\phi(s)
\right\|^2_{L^2({\bf R}^n)} \leq \left( p^{2 \theta} U(t,s) \phi(s) , U(t,s) \phi(s) \right)_{L^2({\bf R}^n)} \leq \tilde{R} \left\| 
\phi(s)
\right\|^2_{L^2({\bf R}^n)}
\end{align*} 
holds. 
\end{Thm} 
As the corollary, we can obtain the following energy stableness property; 
\begin{Thm}[Energy stableness] \label{T2}
Let $ \phi (s) \in H^{\theta}({\bf R}^n)$ satisfies 
\begin{align*}
{a}_1 \left\| 
\phi(s)
\right\|^2_{L^2({\bf R}^n)} \leq \left( H(s)\phi(s) ,  \phi(s) \right)_{L^2({\bf R}^n)} \leq {R}_1 \left\| 
\phi(s)
\right\|^2_{L^2({\bf R}^n)}
\end{align*}
for some constants $0< {a}_1< {R}_1 $. Assume Assumption \ref{Ass1}, \ref{Ass2} and \ref{Ass3}. Then for any $t \in {\bf R}$, there exist $t$-independent constants $0<\tilde{a}_1< \tilde{R}_1$ such that 
\begin{align*}
\tilde{a}_1 \left\| 
\phi(s)
\right\|^2_{L^2({\bf R}^n)} \leq \left( H(t) U(t,s)\phi(s) ,  U(t,s)\phi(s) \right)_{L^2({\bf R}^n)} \leq \tilde{R}_1 \left\| 
\phi(s)
\right\|^2_{L^2({\bf R}^n)} .
\end{align*}
\end{Thm} 
\begin{Cor}[Uniformly stableness in $H^{\theta}({\bf R}^n)$] \label{C1}
Suppose $\phi (s) \in H^{\theta}({\bf R}^n)$. If we assume Assumption \ref{Ass1} and \ref{Ass2}. Then for any $t \in {\bf R}$ there exists a $t-$independent constant $c_m>0$ such that 
\begin{align} \label{7}
\left\| U(t,s) \phi (s) \right\|_{H^{\theta}({\bf R}^n)}^2 \geq c_m  \left\| 
\phi (s)
\right\|^2_{H^{\theta}({\bf R}^n)}
\end{align}
holds. If we assume Assumption \ref{Ass1} and \ref{Ass3}. Then for any $t \in {\bf R}$ there exists a $t-$independent constant $c_M>0$ such that 
\begin{align} \label{7-2}
\left\| U(t,s) \phi (s) \right\|_{H^{\theta}({\bf R}^n)}^2 \leq c_M  \left\| 
\phi (s)
\right\|^2_{H^{\theta}({\bf R}^n)}
\end{align}
holds
\end{Cor}

As an application of this theorem, we shall introduce the Sobolev type inequality; 
\begin{Cor} \label{C2}
Under the assumption \ref{Ass1} and \ref{Ass3}, for all $0 \leq \gamma \leq \theta$ and $\phi(s) \in H^{\gamma} ({\bf R}^n)$, there exists a constant $c>0$ such that 
\begin{align*}
\left\| 
U(t,s)\phi(s)
\right\|_{L^{p}({\bf R}^n)} \leq c \left\| \phi (s) \right\|_{H^{\gamma}({\bf R}^n)}
\end{align*}
holds, where $p$ satisfies 
\begin{align*}
\frac{1}{p} + \frac{\gamma}{n} = \frac12. 
\end{align*} 
and $c$ does not depend on $t$.
\end{Cor}

For $\theta =1$, energy stableness property can be proven for the case where $\left\| V(t, \cdot) \right\|_{L^{\infty} ({\bf R}^n)}$ decays in $t$ or $\left\| V(t, \cdot) \right\|_{L^{\infty} ({\bf R}^n)}$ is sufficiently small or $V(t,\cdot)$ is periodic in time, respectively. However, in the assumption \ref{Ass1}, we do not assume these conditions, and the energy stableness under Assumption \ref{Ass1} -- \ref{Ass3} has not been seen yet, as far as we know. For such potentials, linear scattering theory (in particular asymptotic completeness), Strchartz estimates (see, e.g., Naibo-Stefanov \cite{NS}) and global well-posedness in $L^{\infty}$ for NLS, also have not been proven yet, as far as we know. The energy stableness and Sobolev inequality may be applicable to such studies.

\section{Proof of Theorem}
In this section, we shall prove Theorem \ref{T1}, Corollary \ref{C1} and \ref{C2}. For simplicity, we assume that $\| \cdot \|_2$ denotes $\| \cdot \|_{L^2({\bf R}^n)}$ and $(\cdot, \cdot)$ denotes $(\cdot, \cdot)_{L^2({\bf R}^n)}$. 

Define $F(t)$ as 
\begin{align*}
F(t) = \| |p|^{\theta} u(t) \|_2^2 = (p^{2\theta} u(t),u(t)), 
\end{align*}
where $u(t) = U(t,s) \phi(s)$. Then, by the assumption \eqref{2} (Hardy type inequality), there exists $C_H >0$ such that  
\begin{align*}
F(t) \geq C_{H}({V}(t) u(t), u(t)) =:G(t). 
\end{align*}
Here, by the simple calculation, we have 
\begin{align*}
\frac{d}{dt} F(t) &= \left( i[{V}(t),p^{2\theta}] u(t),u(t)\right),  
\end{align*}
where $[\cdot, \cdot]$ stands for the commutator of operators. On the other hand, 
\begin{align*}
\frac{d}{dt} G(t) &= C_H\left( 
i[p^{2 \theta}/2m , {V}(t)] u(t), u(t)
\right) + C_H (V'(t) u(t), u(t) )\\ &= 
-(C_H/2m) \frac{d}{dt}F(t) + C_H (V'(t) u(t), u(t) )
\end{align*}
Hence we have 
\begin{align} \label{5}
G(t) = C_H \int_s^t (V'(\tau) u(\tau), u(\tau) ) d \tau - \frac{C_H}{2m} \left( F(t) -F(s) \right) + G(s)
\end{align}
Condition $F(t) \geq G(t)$ implies 
\begin{align*}
& \frac{2m + C_H}{ 2m} F(t)  \geq \frac{C_H}{2m} F(s) + G(s) + C_H \int_s^t \left( V'(\tau)u(\tau) , u (\tau) \right)d \tau \\ & \geq \frac{C_H}{4m} F(s) + C_H \left( 
\left( \frac{1}{4m}p^{2 \theta} +V(s) - \int_s^t \left\| V'(\tau, \cdot) \right\|_{L^{\infty} ({\bf R}^n)} d \tau \right) \phi(s) , \phi (s) \right). 
\end{align*}
If we assume Assumption \ref{Ass2}, we notice that there exists a positive constant $C>0$ such that 
\begin{align*}
F(t) \geq C F(s), 
\end{align*}
which implies \eqref{3} holds. On the other hand, if we assume Assumption \ref{Ass3}, then we have   
\begin{align} \nn 
 \frac{C_H}{2m} F(t) & = \frac{C_H}{2m \delta} \left( \delta p^{2 \theta} u(t),u(t) \right) \leq \frac{C_H}{2m \delta}  \left( \left( \frac{1}{2m} p^{2 \theta} + V(t) \right) u(t),u(t) \right) \\ \nn &= \frac{1}{2 m \delta} \left( \frac{C_H}{2m}  F(t) + G(t) \right) \\ & \leq \frac{1}{2m \delta} \left( 
G(s) + \frac{C_H}{2m} F(s) + C_H \int_s^t \left\| V'(\tau, \cdot) \right\|_{L^{\infty} ({\bf R}^n)} d \tau \left\| \phi(s) \right\| ^2_2 \right), \label{6}
\end{align}
which implies \eqref{4} holds. Inequalities \eqref{7} and \eqref{7-2} also hold by \eqref{6} and \eqref{2}. Theorem \ref{T2} can be immediately proven by using following inequalities 
\begin{align*}
(H(t) u(t),u(t) ) &\leq \left( \left( p^{2 \theta}/(2m) + V(t) \right) u(t), u(t)\right)  \\ & \leq (1/2m + 1/C_H) (p^{2 \theta}u(t), u(t) ) 
\end{align*}
and \eqref{9}. 

Finally, we prove Corollary \ref{C2}. By interpolating the followings 
\begin{align*}
\left\| 
|p|^0 U(t,s) \J{p}^{-0}
\right\|_{\SCR{B}(L^2({\bf R}^n))} \leq C_0 
\end{align*}
and 
\begin{align*}
\left\| 
|p|^{\theta} U(t,s) \J{p}^{- \theta}
\right\|_{\SCR{B}(L^2({\bf R}^n))} \leq C_1,  
\end{align*}
we get for some $0 \leq \gamma_0 \leq \theta$ and $(t,s)-$independent constant $C_{\gamma_0} >0$,  
\begin{align*}
\left\| 
|p|^{\gamma_0} U(t,s) \J{p}^{-\gamma_0}
\right\|_{\SCR{B}(L^2({\bf R}^n))} \leq C_{\gamma_0},
\end{align*}
where $\J{\cdot} = (1 + \cdot ^2)^{1/2} $, $C_0 >0$ and $C_1 >0$ are $(t,s)-$independent constants and $\SCR{B} (L^2({\bf R}^n))$ stands for the operator norm on $L^2({\bf R}^n)$. That provides for a pair $(p,\gamma )$ with $1/p + \gamma /n =1/2$,
\begin{align*}
\left\| 
U(t,s) \phi (s)
\right\|_{L^{p} ({\bf R}^n)} & \leq  C \left\| |p|^{\gamma} U(t,s) \phi (s) \right\|_2  \\ & \leq 
C  \left\| |p|^{\gamma} U(t,s) \J{p}^{-\gamma} \right\|_{\SCR{B}(L^2({\bf R}^n))} \left\| \J{p}^{\gamma} \phi (s) \right\|_2, 
\end{align*}
and which is the desired result, where we use the Gagliardo-Nirenberg inequality.

\end{document}